\documentclass[12pt]{article}
\usepackage{amsmath}
\usepackage{epstopdf}
\usepackage{graphicx}

\makeatletter

\@addtoreset{figure}{section}
\def\thefigure{\thesection.\@arabic\c@figure}
\def\fps@figure{h, t}
\@addtoreset{table}{bsection}
\def\thetable{\thesection.\@arabic\c@table}
\def\fps@table{h, t}
\@addtoreset{equation}{section}

\makeatother

\begin{document}

\title{Magneto-optical granulometry: on the determination of the statistics of magnetically induced particle chains in concentrated ferrofluids from linear dichroism experiments}

\author{V. Socoliuc\thanks {corresponding author: vsocoliuc@acad-tim.tm.edu.ro, vsocoliuc@gmail.com} \thanks{Romanian Academy - Timisoara Branch, Center for Fundamental and Advanced Technical Research, Laboratory of Magnetic Fluids, Bv.M. Viteazu 24, Timisoara RO-300223, Romania}  \space  and L.B. Popescu\thanks{Institute for Space Sciences, 409 Atomistilor, Magurele RO-077125, Romania}}

\date{\today}
\maketitle

\begin{abstract}
An analytical theoretical model for the influence of the magnetically induced nanoparticle chaining on the linear dichroism in ferrofluids was developed. The model is based on a statistical theory for magnetic nanoparticle chaining in ferrofluids. Together with appropriate experimental approach and data processing strategy, the model grounds a magneto-optical granulometry method able to determine the magnetic field dependence of the statistics of magnetically induced particle chains in concentrated ferrofluids.
\end{abstract}

\section{Introduction\label{intro}}
Ferrofluids,  i.e. magnetic fluids or magnetic nanofluids, are stable colloidal dispersions of magnetic monodomain nanoparticles dispersed in a carrier fluid \cite{Rozi, BerkoBashto}. Magneto-optical effects have been known to be useful tools for micro-structural investigations of ferrofluids. Experimental data \cite{Take,Socoliuc1999} show nontrivial dependence of the specific dichroism (the dichroism divided by the nanoparticle volume fraction) on the ferrofluid nanoparticle volume fraction. This contradicts the dichroism models for ferrofluids with noninteracting nanoparticles \cite{Take,Hasmonay} which predict that the specific dichroism is independent on the ferrofluid nanoparticle volume fraction. Therefore, magnetically induced particle chaining should be considered as one of the main causes of the linear dichroism and birefringence in concentrated ferrofluids \cite{Take, Xu, Socoliuc1999, IvaKanto}.

In this paper we develop a dichroism model for ferrofluids with magnetically induced nanoparticle chains.  The nanoparticle chaining is accounted by using the statistical model developed by Mendelev and Ivanov \cite{MendeIva}. The resulting analytical model will be used to process dichroism experimental data from ref.\cite{Socoliuc1999} by means of nonlinear regression, in order to determine the magnetic field dependence of the particle chains' statistics in a ferrofluid with polydisperse magnetite nanoparticles.

\section{Materials and methods\label{mathmet}}

The dichroism experiments were carried on a transformer oil based ferrofluid with magnetite nanoparticles sterically stabilized with oleic acid molecules. The ferrofluid was synthesized by means of the chemical coprecipitation method \cite{Bica} at {\it The Laboratory of Magnetic Fluids, Center for Fundamental and Advanced Technical Research, Romanian Academy - Timisoara Branch}. As will be shown below, the ferrofluid contains a small but not negligible amount of spontaneous nanoparticle clusters. Therefore, we shall further designate by the term "colloidal particles" the ensemble of nanoparticles and spontaneous nanoparticle clusters that constitutes the sol-liquid colloidal state of the ferrofluid. Two ferrofluid samples were investigated: a highly concentrated sample ($S_1$) with 9.3\% nanoparticle volume fraction, and a 40 times diluted sample ($S_2$)  with 0.23\% nanoparticle volume fraction. $S_1$ was vigorously stirred prior to dilution in order that both samples contain colloidal particles with the same size and morphology statistics. For details regarding sample synthesis and dilution see ref.\cite{Socoliuc1999}  

TEM and magneto granulometry \cite{IvaMgngran} investigations were done in order to determine the statistics of the nanoparticles and colloidal particles in the ferrofluid. Static Light Scattering (SLS) \cite{SocoliucVekasTurcu} investigations were made in order to test the ferrofluid samples' susceptibility to undergo magnetically induced phase separation. The dichroism of the samples was measured by means of the transmission method. Both ferrofluid samples were measured at 22$^\circ$C and were contained in a quartz cell with 0.1mm optical path. Details on the experimental setup and procedure can be found in refs. \cite{Socoliuc1999} and \cite{SocoliucPopescu2012}.

\section{Theoretical basis\label{theor}}

The dichroism of a ferrofluid with polydisperse and ellipsoidal N\'{e}el nanoparticles,  i.e. soft magnetic nanoparticles whose magnetic moment is coupled to the symmetry axis by the shape anisotropy energy, was derived in an analytical form by Hasmonay and coworkers \cite{Hasmonay}:
\begin{equation}
\begin{split}
 \Delta
&
Im \left(n\right)=
\Phi\frac{n_1}{2} 
\int_{0}^{\infty} \Delta Im(\chi)\left(1-\frac{3L\left(\xi\right)}{\xi}\right) \times 
\\ &
 \times \frac{1}{4}\left( \frac{6}{\sqrt{\pi}}\cdot \frac{e^{\eta}}{\sqrt{\eta}\cdot Erfi\left(\sqrt{\eta}\right)}-\frac{3}{\eta}-2\right)\cdot f(D) \cdot dD .
\end{split}
\label{eq1}
\end{equation}
 $\Phi$ is the volume fraction of the nanoparticles in the ferrofluid. $n_1$ is the refractive index of the ferrofluid carrier ($n_1$=1.457 for the transformer oil, measured by means of refractometry). $\bar\chi$  is the dielectric susceptibility tensor of the ellipsoidal particles and $\Delta Im(\chi) = Im(\chi_{z})-Im(\chi_{x})$ is a function on the average aspect ratio $<r>$ and the complex refractive index of magnetite \cite{Xu}. $L\left(x\right)=coth\left(x\right)-1/x$ is the Langevin function \cite{Rozi}. $\xi=\xi\left(H,D\right)$ and $\eta=\eta\left(D,<r>\right)$ are the magnetic dipole-field and magnetic anisotropy energies of the free coloidal particles, normalized to the thermal energy $k_{B}T$ \cite{Hasmonay}. $f(D)$ is the log-normal probability distribution function of the nanoparticle diameter, and depends on two parameters: the median diameter $D_0$ and the distribution width $\sigma$. Thus, the magnetically induced dichroism is a function on the magnetic field intensity $H$, nanoparticle volume fraction $\Phi$, nanoparticle size statistics $f(D)$ and nanoparticle average aspect ratio $<r>$. In the case of Brown particles, when the magnetic moment is fixed along the shape anisotropy axis,  $\eta\rightarrow \infty$ and the third factor in the integral of eq.\ref{eq1} equals unity. Thus eq.\ref{eq1} becomes the well known expression of the dichroism for ferrofluids with ellipsoidal Brown nanoparticles \cite{BerkoBashto, Take}.

It is important to observe that eq.\ref{eq1} predicts that the specific dichroism $\Delta Im(n) / \Phi$ is independent on the nanoparticle volume fraction in a ferrofluid where no magnetically induced particle clustering occurs.

The dichroism of a ferrofluid with magnetically induced nanoparticle chains can be obtained from eq.\ref{eq1} by replacing the integral over the nanoparticle diameter with the summation over the number of particles per chain \cite{SocoliucPopescu2012}. We shall assume monodispersed spherical nanoparticles and that the resultant magnetic moment of the chain is parallel to its anisotropy axis. If the chains behave like Brown particles, one may drop the third term in the integral of eq.\ref{eq1}, so the dichroism expression is: 
\begin{equation}
\begin{split}
 \Delta Im\left(n\right)=&\Phi\frac{n_1}{2} \times \\ &  \times\sum_{p=2}^{\infty}w_p\cdot\Delta Im(\chi_p)\cdot \left(1-\frac{3L\left(\xi_p\right)}{\xi_p}\right).
\end{split}
\label{eq2}
\end{equation}
The summation starts at p=2 because the spherical monomers (p=1) do not contribute to the dichroism ($\Delta Im(\chi_{sphere})=0$ \cite{Xu}).  $\chi_p$ is the dielectric susceptibility of the p-particle chain given in detail in ref. \cite{Xu}. For the p-particle chain: $\xi_p=p\cdot\xi$.  $w_p=w_p\left(H,\Phi,D\right)$ is the volume weight from $\Phi$ of the p-particle chains ($\sum_{p=1}^{\infty}w_p=1$), and its analytical expression for monodispersed spherical nanoparticles is given in detail in ref.\cite{MendeIva}.

Again, it is important to stress that eq.\eqref{eq2} predicts that the specific dichroism $\Delta Im(n) / \Phi$ in ferrofluids with magnetically induced nanoparticle chains depends on the nanoparticle volume fraction via the chains' weight  $w_p\left(H,\Phi,D\right)$.

Eqs.\ref{eq1} and \ref{eq2} will serve the basis for the development in the following section of a dichroism model in concentrated ferrofluids, where the contribution of both ellipsoidal nanoparticles and magnetically induced nanoparticle chains must be taken into account in order to explain the experimental data.

\section{Results and discussion\label{resdis}}

No on-shelf sedimentation of the ferrofluid samples was observed even after months of rest. From SLS experiments it was found that both ferrofluid samples do not undergo magnetically induced phase condensation in the field range 0 - 300 kA/m and temperature range 15 - 80$^\circ$C.

\begin{figure}
\centerline{\includegraphics{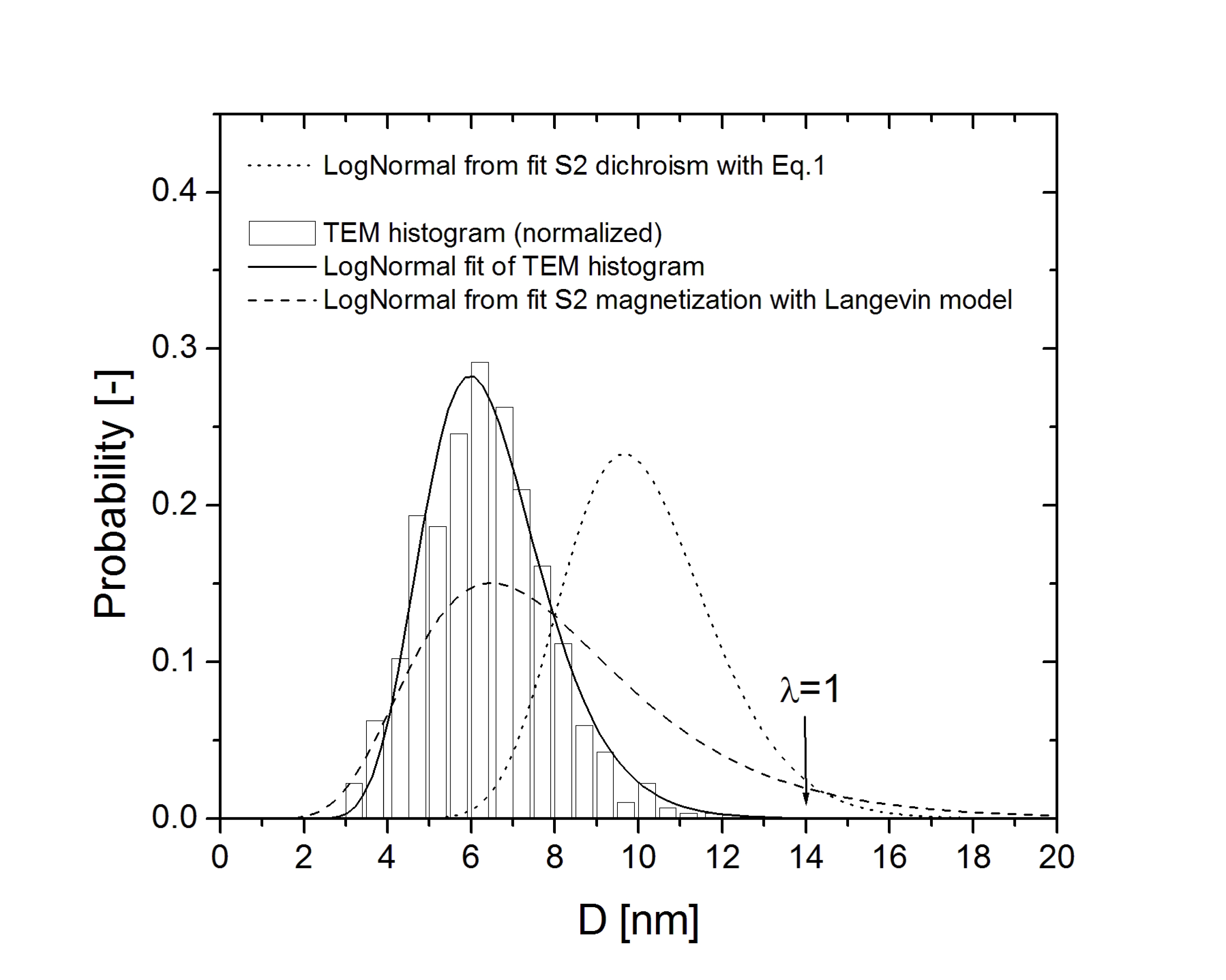}}
\caption{\label{fig1} Granulometry results from TEM, magnetization and dichroism.}
\end{figure}

The average physical diameter $D$ and aspect ratio $r$ of the nanoparticles was determined from TEM investigations on a sample of more than 2000 nanoparticles: $<D>=6.3 \pm 0.1 nm$ and $<r>=1.24 \pm 0.01$. The correlation coefficient between diameter and ellipticity is 0.018, therefore they can be considered statistically independent. The size statistics of the nanoparticles is well described by a log-normal distribution (fig.\ref{fig1}). Due to nanoparticle clustering in the process of the grid preparation, it is impossible to determine the size statistics of the spontaneous particle clusters from TEM.

The measured field $H$ dependence of the diluted sample $S_2$ magnetization $M(H)$ was processed by means of magneto-granulometry (i.e. fitting the measured $M(H)$ with the log-normal integral Langevin model) assuming negligible magnetically induced clustering in the 0.23\% highly diluted $S_2$ sample. For details on magneto-granulometry see ref.\cite{IvaMgngran}. The dashed line in fig.\ref{fig1} plots the log-normal distribution ($D_0 = 7.5 \pm 0.1 nm$ and $\sigma = 0.38 \pm 0.012$) of the physical diameter as obtained from magneto granulometry. One can notice that the magneto-granulometry distribution is wider than the TEM one, revealing the existence of spontaneous nanoparticle clusters in the structure of the ferrofluid, most likely due to either incomplete surfactant covering or bridge interactions among surfactant molecules. Similar observations were confirmed by rheogical investigations in ref.\cite{Resiga}. Thus, the  magneto-granulometry distribution indicates the possibility of magnetically induced clustering of the colloidal particles, since the tail of the distribution extends beyond the diameter value for which the interaction parameter of the magnetic dipole-dipole interaction $\lambda$ equals unity.

\begin{equation}
\lambda \stackrel{\mbox{def}}=\frac{1}{2}\cdot \frac{U_{m,d-d,MAX}}{k_{B}T}=1.
\label{eq3}
\end{equation}
Because the Langevin model used for magneto-granulometry assumes monodomain magnetic nanoparticles, the diameter of the spontaneous clusters has the significance of a magnetic effective diameter.

\begin{figure}
\centerline{\includegraphics{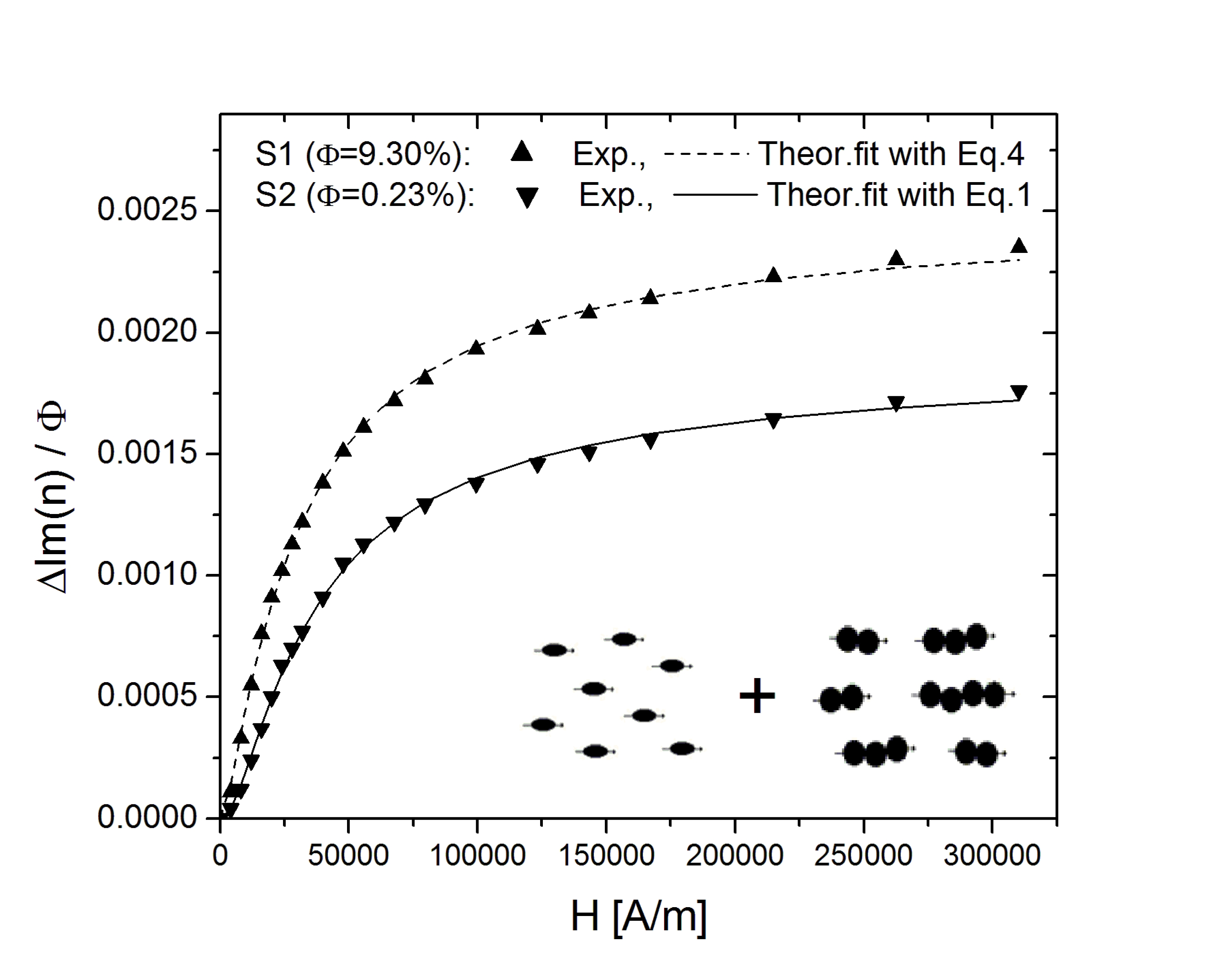}}
\caption{\label{fig2}Magnetically induced dichroism in samples $S_1$ and $S_2$: experiment (dots) and theoretical fits (lines). }
\end{figure}

The field intensity dependence of the magnetically induced dichroism in samples $S_1$ and $S_2$ are presented in fig.\ref{fig2}. $\Delta Im\left(n\right)/\Phi\mid_{\text{exp}} =\left( Im(n_{\parallel,exp})-Im(n_{\perp,exp}) \right)/\Phi$ is the specific, i.e. reduced measured dichroism, the difference between the imaginary refraction indexes for light polarized parallel and perpendicular to the magnetic field direction, divided by the nanoparticle volume fraction  ${\Phi}$ of the ferrofluid sample. The most important observation is that the specific dichroism of the concentrated sample $S_1$ is about 25\% larger than that of the highly diluted sample $S_2$. As shown in the previous section, should only the rotation of the colloidal particles were responsible for the magnetically induced dichroism, the specific dichroism were independent on the sample's nanoparticle volume fraction. Therefore, since the samples show no magnetically induced phase condensation and the magnetically induced chain formation is negligible in the highly diluted sample $S_2$ \cite{MendeIva}, it is plausible to assume that the difference between the measured magnetically induced dichroism of the samples is due to the magnetically induced particle chaining in the concentrated sample $S_1$.

The dichroism experimental data was processed by means of nonlinear regression. The nonlinear regression was done in {\it Mathematica\textsuperscript{\textregistered}} using the Levenberg-Marquardt method.

Assuming that the magnetically induced particle chaining is negligible in sample $S_2$, its dichroism curve was fitted with eq.\ref{eq1}. The fit curve with $R^2 = 0.9986$ is presented in fig.\ref{fig2}. The resulting log-normal size distribution ($D_0 = 10.0 \pm 0.2 nm$ and $\sigma = 0.17 \pm 0.2$) is plotted in fig.\ref{fig1} (dot line) and the average ellipticity is $<r> = 1.56 \pm 0.02$. The size distribution is poorly expressed in the small diameter region because the maximum of the magnetic field intensity in the experimental setup is not enough to saturate the dichroism. Compared with the TEM results for the size distribution and ellipticity, the dichroism granulometry reveals spontaneous particle clustering, mostly dimers since the value of the ellipticity is significantly larger than the TEM value ($<r>_{TEM}=1.24 \pm 0.01$), but smaller than 2. The spontaneous particle clustering leads to the formation of colloidal particles with effective diameter large enough ($\lambda >1$) to undergo magnetically induced chaining, as also concluded from the magneto-granulometry analysis (fig.\ref{fig1}).

Following the above results, two basic assumptions will be made for the purpose of fitting the dichroism curve of the concentrated sample $S_1$: {\bf {i.}} the statistics of the colloidal particles in $S_1$ will be reduced to a bidisperse distribution of small ($\lambda <1$) and large particles respectively ($\lambda >1$), and {\bf {ii.}} the magnetically induced dichroism of sample $S_1$ is the sum of the dichroism caused by the small colloidal particles and the large colloidal particle chains respectively. As a result, the difference between specific dichroisms of samples $S_1$ and $S_2$ ($\delta\Delta Im(n) /\Phi_1=\Delta Im(n)/\Phi\mid_{S_1}-\Delta Im(n)/\Phi\mid_{S_2}$) is solely due to the fraction of large colloidal particles in $S_1$. Consequently, combining eqs. \ref{eq1} and \ref{eq2}, the theoretical expression for $\delta \Delta Im(n)$ is:
\begin{equation}
\begin{split}
&\delta \Delta Im(n)=\Phi_l\frac{n_1}{8}w_1\cdot\Delta Im(\chi_1)\cdot\left(1-\frac{3L\left(\xi_1\right)}{\xi_1}\right) \times
 \left( \frac{6}{\sqrt{\pi}}\cdot \frac{e^{\eta}}{\sqrt{\eta}\cdot Erfi\left(\sqrt{\eta}\right)}-\frac{3}{\eta}-2\right)+
\\&
+\Phi_l \frac{n_1}{2}\sum_{p=2}^{\infty}w_p\cdot\Delta Im(\chi_p)\cdot \left(1-\frac{3L\left(\xi_p\right)}{\xi_p}\right),
\end{split}
\label{eq4}
\end{equation}
where $\Phi_l$ and $D_l$ are the volume fraction and diameter of the large colloidal particles. The first term in the r.h.s. of eq.\ref{eq4} is the dichroism of unaggregated large colloidal particles with weight $w_1(H, \Phi_l, D_l)$, assumed to be N\'{e}el ellipsoidal particles with the average ellipticity $<r>=1.56$ determined from the fit for the diluted sample $S_2$. The second term in the r.h.s. of eq.\ref{eq4} is the dichroism caused by the p-particle chains, where the colloidal particles will be assumed spherical. eq.\ref{eq4} is an analytical expression with variable H, and parameters $\Phi_l$ and $D_l$ .

The experimental magnetic field intensity dependence for $\delta \Delta Im(n)$ was calculated from the data in fig.\ref{fig2} and was fitted with eq.\ref{eq4}. The resulting fit parameter values are: $\Phi_l=0.67\pm0.03 \% $ and $D_l=15.3\pm0.2nm$. The fit curve is presented in fig.\ref{fig2} after being incremented with the fit curve of the diluted sample $S_2$ The $R^2 = 0.9989$ of the incremented theoretical curve shows a good fit. The volume fraction of the large colloidal particles in $S_1$ is $\Phi_l=0.67\%$, which is $7.2\%$ volume from the total magnetic nanoparticle dispersed in the sample. The resulting diameter of the large colloidal particles is greater than ~14nm, the diameter for which $\lambda=1$, thus confirming the possibility for the magnetically induced chain formation.

\begin{figure}
\centerline{\includegraphics{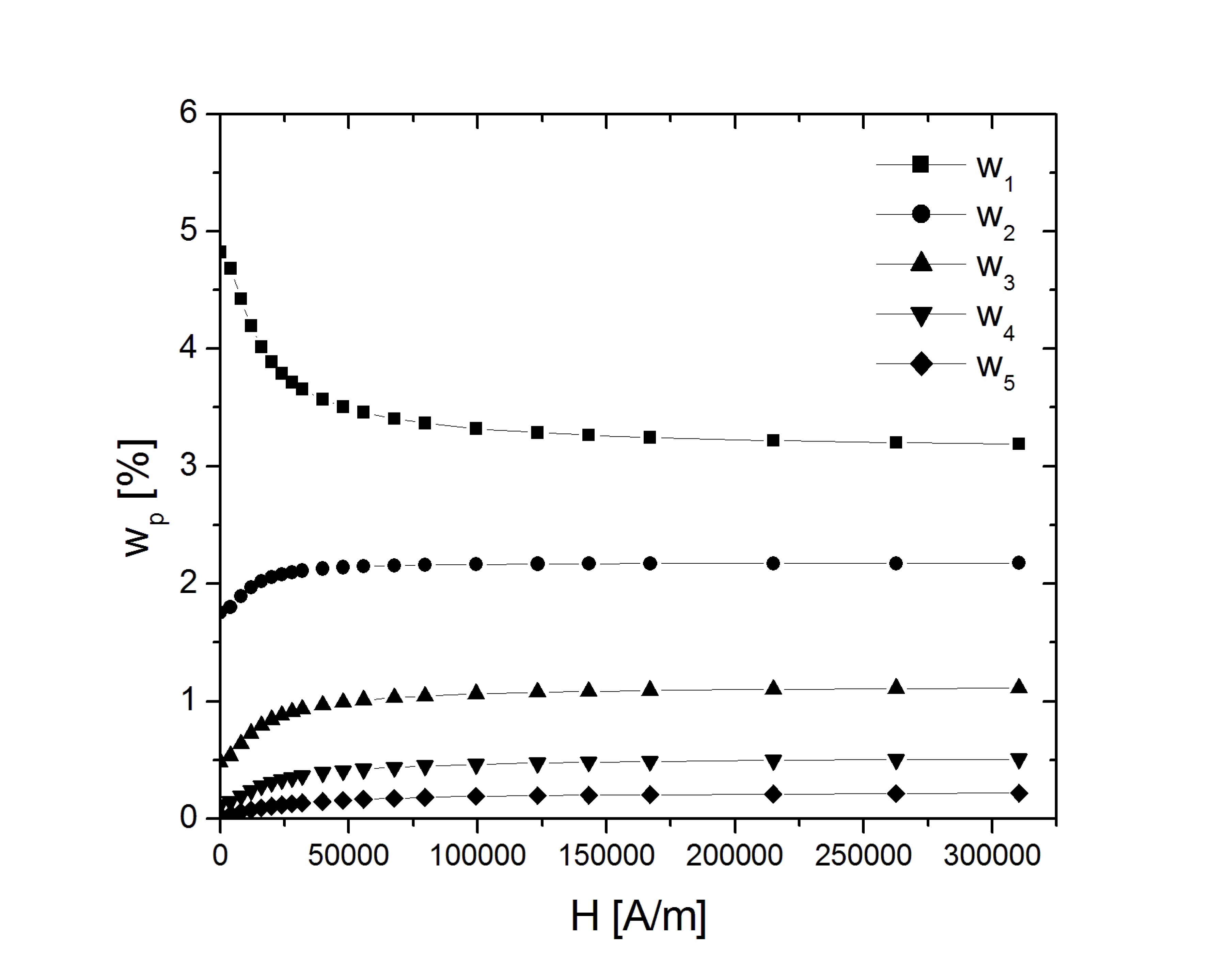}}
\caption{\label{fig3} Magnetic field intensity dependence of the weights of free particle (p=1) and particle chains (p=2...5).}
\end{figure}

\begin{figure}
\centerline{\includegraphics{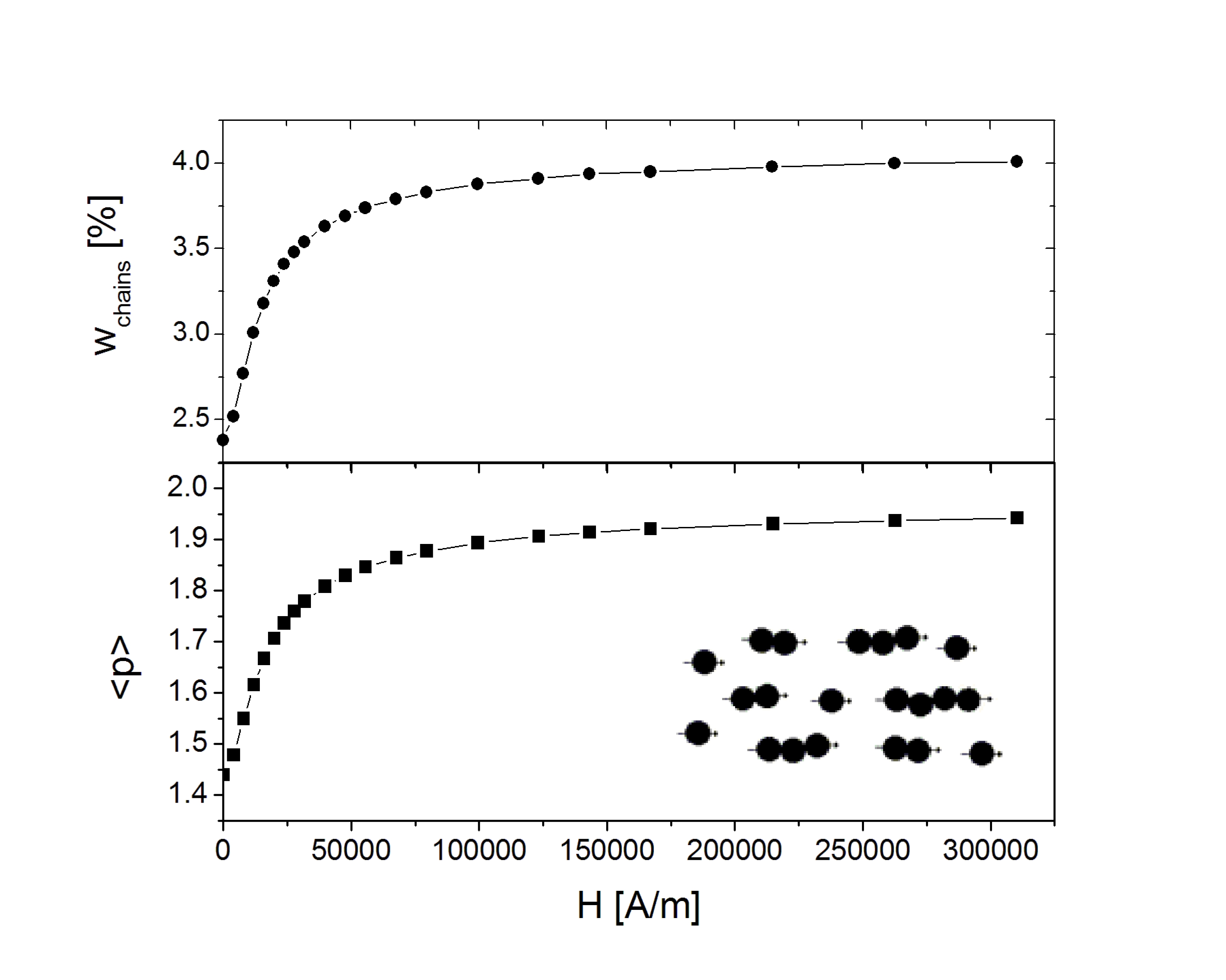}}
\caption{\label{fig4} Magnetic field intensity dependence of the average number of particles per chain ($<p>$) and the total volume weight of particle chains ($w_{chains}$).}
\end{figure}

With the fit values for $\Phi_l$ and $D_l$ the magnetic field intensity dependence of the weights $w_p$ was calculated using the theoretical model of Mendeleev and Ivanov \cite{MendeIva}, and plotted in fig.\ref{fig3} for p=1...5. The field dependence of the average number of particles per chain $<p>$ and the volume percentage of all particle chains (p=2...5) from the total magnetic nanoparticle dispersed in the sample are also plotted in fig.\ref{fig4}.

One can notice that, with increasing field intensity H, the chain weights increase in the detriment of the weight of free particles $w_1$. The weight of the chains is diminishing with increasing $p$. As the field intensity grows, the chaining process saturates because the dipole-field interaction overcomes the dipole-dipole interaction. At saturation, the average number of particles per chain is $<p> \approx 2$, while the total chains' volume wight is $w_{chains} \approx 4\%$. Although small, a 4\% weight of magnetically induced chains accounts for 25\% from the specific dichroism of the concentrated sample $S_1$. This observation is important for at least two reasons:  {\bf {i.}} it stresses that it is compulsory to consider the magnetically induced particle chaining in the analysis of the dichroism of concentrated ferrofluids, and  {\bf {ii.}} it shows that dichroism experiments are a highly sensitive tool for the investigation of magnetically induced particle chaining in ferrofluids.

In zero field, the theory predicts that there is a significant amount of magnetically induced particle chains, due to the fact that for 15.3nm diameter particles the magnetic dipole-dipole interaction parameter is $\lambda=2.1>1$. This is because, in order to keep the theoretical model analytically evaluable, we approximated the magnetic spontaneous clusters as monodomain particles. A rigorous model, albeit only numerically evaluable, will predict no chain weights in the absence of the external magnetic field because spontaneous clusters of nanoparticles less than 10nm diameter will have no resultant magnetic moment in the absence of the field.
\bigskip

\section{Conclusions}

The proposed theoretical model for the magnetically induced dichroism enables the development of a method for the determination of the magnetically induced particle chain statistics in concentrated ferrofluids, i.e. {\it magneto-optical granulometry}. Shifting from the imaginary to the real part of the ferrofluid refraction index, the method can be straightforward applied to birefringence experiments. The application of the method to dichroism experimental data of a transformer oil based ferrofluid with $\Phi=9.3\%$ solid volume fraction, revealed the increase of the average number of particles per chain up to $<p> \approx 2$ and of the chains volume fraction up to $\Phi_{chains} \approx 0.4\%$, in a 300 kA/m magnetic field. Thus, the magneto-optical granulometry turns linear dichroism and birefringence experiments into a very sensitive tool for the investigation of the structure of ferrofluids, with numerous applications in both fundamental and applied research.
\newline

\section*{Acknowledgements}
V. Socoliuc acknowledges the financial support of {\it ARFT-CCTFA-LLM 2013-2015} research program. L.B. Popescu acknowledges the financial support of {\it Programul NUCLEU LAPLAS 3} PN 09.39.06.04 and PN 09.39.06.08. The authors are grateful to Oana Marinica from {\it Politehnica Unversity of Timisoara}  for the VSM measurements. This work is dedicated to the memory of Dr. Doina Bica (1952 - 2008).

\bibliographystyle{new}

\begin{thebibliography}{300}

\bibitem{Rozi}
R. E. Rosensweig, Ferrohydrodynamics, Cambridge Univ. Press, Cambridge, 1985.

\bibitem{BerkoBashto}
B. Berkovski, V. Bashtovoy (Eds.), Magnetic Fluids and Applications Handbook, Begell House, New York, 1996.

\bibitem{Take}
S. Taketomi, M. Ukita, M. Misukami, H. Miajima, S. Chikazumi, J. Phys. Soc. Japan 56 (1987) 3362.

\bibitem{Socoliuc1999}
V. Socoliuc, J.Magn.Magn.Mater 207 (1999) 146.

\bibitem{Hasmonay}
E. Hasmonay, E.Dubois, J.-C.Bacri, R.Perzynski, Yu.L.Raikher, V.I.Stepanov, European Physical Journal B 5 (1998) 859–867.

\bibitem{Xu}
M. Xu, P.J. Ridler, J. Appl. Phys. 82 (1997) 326.

\bibitem{IvaKanto}
A.O. Ivanov and S. Kantorovich, Phys.Rev.E 70 (2004) 021401.

\bibitem{MendeIva}
V.S. Mendelev, A.O. Ivanov, Phys. Rev. E 70 (2004) 051502.

\bibitem{Bica}
D. Bica, Rom. Rep. Phys. 47 (1995) 265.

\bibitem{IvaMgngran}
A. O. Ivanov, S. S. Kantorovich, E. N. Reznikov, C. Holm, A. F. Pshenichnikov, A. V. Lebedev, A. Chremos and P. J. Camp, Phys. Rev. E: Stat., Nonlinear, Soft Matter Phys., 75 (2007) 061405.

\bibitem{SocoliucVekasTurcu}
V. Socoliuc, L. Vekas, R. Turcu, Soft Matter 9 (2013) 9.

\bibitem{SocoliucPopescu2012}
V. Socoliuc, L.B. Popescu, J.Magn.Magn.Mater 324 (2012) 113.

\bibitem{Resiga}
D. Susan-Resiga, V. Socoliuc, T. Boros, T. Borbath, O. Marinica, A. Han, L. Vekas, J.Coll.Int.Sci. 373 (2012) 110.

\end{thebibliography}

\end{document}